\documentclass{phc-proc4-auth}

\usepackage{graphicx}

\usepackage{amssymb}

\usepackage{amsmath}

\begin{document}

\begin{frontmatter}



\title{Superconductor-insulator transition induced by pressure
within the X-boson approach
\thanksref{CNPq}
}
\thanks[CNPq]{We acknowledge Prof.  M. E. Foglio (Unicamp) for helpful discussions
and the financial support from the Rio de Janeiro
State Research Foundation (FAPERJ)
and National Research Council (CNPq).}


\author[IfUFF]{Lizardo H. C. M. Nunes},
\ead{lizardo@if.uff.br}
\author[IfUFF]{M. S. Figueira},
\author[IfUFF]{E. V. L. de Melllo}

\address[IfUFF]
{Instituto de F\'{\i}sica da
Universidade Federal Fluminense\\
Av. Litor\^{a}nea s/n, 24210-340 Niter\'oi, Rio de Janeiro, Brasil.
C.P.100.093}

\begin{abstract}
The pressure induced superconducting phase diagram
is calculated for an extension of the periodic
Anderson model (PAM) in the $ U=\infty $ limit
taking into account the effect of a nearest
neighbor attractive interaction between $ f $-electrons.
We analyze the role of the
chemical potential compared to several plots of
the the $ f $-band density of states and
we also found a superconductor-insulator transition
induced by pressure when
the chemical potential crosses the hybridization gap.
\end{abstract}

\end{frontmatter}


\section{Introduction}\label{int}

Recently, we have used the X-boson method
\cite{X-boson} to
obtain the superconducting phase diagram of the
PAM in the $ U \rightarrow \infty $,
where $ U $ is the on-site
Coulomb repulsion, plus
a nearest neighbor attractive interaction between
the localized $ f $-electrons.
As charge carriers are added
to the system \cite{Nunes2002}
superconductivity was found both for
configurations where the system presented intermediate valence (IV) and heavy fermion (HF)
behavior, as experimentally observed\cite{portugueses}.
Moreover, we have recovered the three
characteristic regimes of the PAM:
Kondo, IV and  magnetic; which cannot be found for the
same model by the slave-boson treatment\cite{portugueses},
since it breaks down
when the $ f $-occupation number $ N_{ f } \rightarrow 1 $.
In this paper we use the same approach to study the pressure induced superconducting phase diagram of the heavy fermion systems.

\section{The Method} \label{theMethod}

In our approach
we consider the energy of the localized $ f $-electrons
site independent,
${ E }_{f,j,\sigma}={ E }_{f}$,
and impose a constant hybridization,
$V=V_{j,\sigma,\mathbf{k}}$,
\begin{eqnarray} \label{model}
H  & = & \sum_{ {\bf k}, \sigma } {\epsilon}_{ {\bf k}, \sigma}
c^{\dagger}_{ {\bf k}, \sigma} c_{ {\bf k}, \sigma} + \sum_{ {\bf
k}, \sigma} \tilde{{E}}_{f} X_{{\bf k},\sigma \sigma }
\nonumber \\
& &
+ \sum_{{\bf k}, \sigma } V (X_{\bf
{k},0\sigma}^{\dagger}c_{{\bf k},\sigma} +  c^{\dagger}_{\bf {k},
\sigma}X_{\bf {k}, 0\sigma } )
\nonumber \\
& &
+ \sum_{ {\bf k}, {\bf k'} } W_{ {\bf k}, {\bf k'} }
b^{\dagger}_{ {\bf k} } b_{ {\bf k'} } + N_{s} \Lambda \left( R - 1 \right) \, ,
\end{eqnarray}
where the operator
$
b^{\dagger}_{{\bf k } } = X_{\bf {k},0\sigma}^{\dagger} X_{-\bf
{k},0\overline{\sigma}}^{\dagger}
$, creates Cooper pairs.
The matrix element
$ W_{\bf {k},\bf {k'}} $ is
the superconducting coupling.
Since we only consider
an isotropic $ s $-wave
superconducting gap,
$ W_{\bf {k},\bf {k'}} $
is reduced to the well-known BCS interaction.
Also, the operator
$ X_{ {\bf k } ,\sigma \sigma'} $
is the Hubbard operator.
The variational parameter
$R=1-\sum_{\sigma}<X_{\sigma\sigma}>$,
introduced by the X-boson approach,
modifies the approximate GF so that
it minimizes an adequate thermodynamic potential while being
forced to satisfy the ``completeness'' relation
$ n_{0}+n_{\sigma}+n_{\overline{\sigma}}=1 \label{Eq.2} $,
where $ n_{j,a}=<X_{j,aa}> $.
Within the X-boson method
the Lagrange multiplier $ \Lambda $
is chosen so that the ``completeness'' is
satisfied and the energy of the $ f $-electrons are
renormalized by
$ \tilde{{E}}_{f} = E_{ f } + \Lambda $.


At $ T = T_{ c } $
the Green's functions (GFs) yield to the
result found in the chain approximation (CHA)
of the PAM,
\cite{X-boson}
but now with the localized
energy levels $ E_{ f } $ and $ D_{\sigma} = R + n_{\sigma } $
renormalized,
and the self-consistent equation for $ T_{ c } $ is \cite{Nunes2002}
\begin{equation}\label{gapi}
1  = - \frac{ W }{ \beta_{c} } \sum_{ n, {\bf k }}
\left|
{ \mathcal{G } }^{ f f }_{\sigma }( { -\bf k }, \omega_{ n } )
\right|_{ T = T_{ c } }^{2}
\, ,
\end{equation}
which is solved
constrained to the ``completeness''
relation and for a constant conduction
density of states,
$ \rho( \epsilon_{\bf k }) = 1 / 2D $,
only defined for the interval
$  -D \le \epsilon_{\bf k } - \mu \le D $.
The pressure induced  phase diagram is obtained
following a scheme proposed
by Bernhard and Lacroix\cite{Lacroix99}
assuming that the ratio $ V / D $ is constant.
Indeed, one expects that the local parameters
$ E_{ f } $ and $ U $ do not change with pressure,
while $ D $ and $ V $ should increase with pressure.
The total electron number
$ N_{ t } = N_{ f } + N_{ c } $,
is kept constant and the
chemical potential is calculated
self-consistently while
the electrons are allowed
to transfer between bands.

\section{Results and Discussion} \label{results}

We choose the parameters to characterize
a typical Kondo regime. We set $ E_{f} = -0.15 $
below the chemical potential potential $ \mu = 0 $
and $ W $ corresponds to $ 5 \% $
of the bandwidth $ 2D $.
\begin{figure}[htb]
\centerline
{
\includegraphics[
angle=90,
width=0.385\textwidth]
{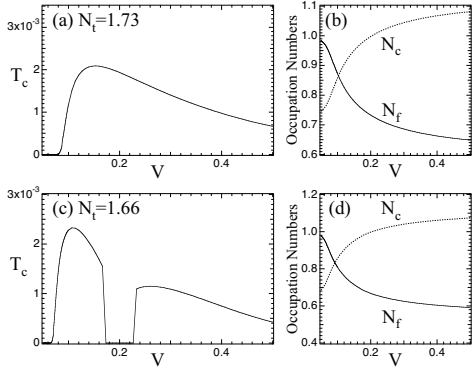}
}
\caption
{
(a) The pressure induced superconducting
phase diagram for $ V/D = 0.2 $ and $ N_{ t } = 1.73 $.
(b) The onset shows the occupation numbers for
the $ f $-electrons $ N_{ f } $
and the conducting electrons $ N_{ c } $
as a function of the hybridization $ V $.
(c)-(d) The same as in (a)-(b) but for $ N_{ t } =1.66 $.
}
\label{FigTcxV}
\end{figure}
Within the X-boson approach and for every
set of parameters used,
as $ N_{ t } $  is kept constant and
$ V $ increases, $ N_{ c } $ also increases.
Hence, electrons
are removed from the $ f $-band
and placed in the conducting band,
as can be seen in Figs. \ref{FigTcxV}.b and
\ref{FigTcxV}.d. Since Cooper pairing
occurs only between the $ f $ electrons,
as $ N_{ f } $ decreases, $ T_{ c} $ also
decreases.
\begin{figure}[htb]
\centerline
{
\includegraphics[
angle=90,
width=0.411\textwidth]
{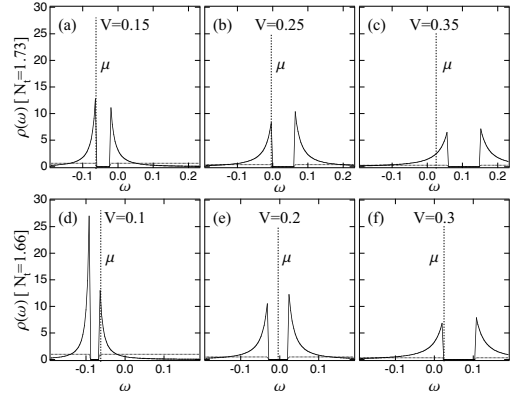}
}
\caption{
Plots of the $ f $-band density of states
$ \rho_{ f  }(\omega) $ (filled line) and the
$ c $-band density of sates $ \rho_{c }(\omega) $ (short dashed lines)
for (a)-(c) $ N_{ t } = 1.73 $ and
(d)-(f) $ N_{ t } = 1.73 $ for $ V / D = 0.2 $
compared to
the values of the chemical potential
$ \mu $ (vertical dashed line).
}
\label{FigDOS}
\end{figure}
Moreover, our results
for the superconducting phase diagram can
be inferred from the behavior of the
chemical potential compared to several plots of
the the $ f $-band density of states.
Indeed, our previous results \cite{Nunes2002}
show that the existence of superconductivity
is constrained to a region where the
$ f $-band densities of states
$ \rho_{f}( \mu ) $ at $ \omega = \mu $ is sufficiently high
while high $ \rho_{f}( \mu )$ usually
implicates higher values of $ T_{c} $.
For low $ V $ the $ f $-band density of states
is located in a narrower region of the bandwidth, since
the  electrons are more localized, the two
peaks of the density of states are sharp and the distance between
them is small, as $ V $ increases,
the $ f $-band density of states takes up smaller values
and the distance between peaks
are broadened, as can be seen in Fig. \ref{FigDOS}.
Hence, for $ N_{ t } = 1.73 $ and based on the
arguments above,
the function $ T_{c}( V ) $
have a single maximum.
For $ N_{ t } = 1.66 $ superconductivity is suppressed
when $\mu $ crosses the hybridization gap and
the system presents a superconductor-insulator transition
induced by pressure. Recently the same kind of transition was experimentally observed for the spinel compound
CuRh$_{2}�$S$_{4}$\cite{Masakazu02}.



\end{document}